\documentstyle[12pt]{article}

 \newcommand{\be}{\begin{equation}}
 \newcommand{\ee}{\end{equation}}
 \newcommand{\ba}{\begin{eqnarray}}
 \newcommand{\ea}{\end{eqnarray}}

\newcommand{\lef}{\left}

\newcommand{\ri}{\right}

\newcommand{\fr}{\frac}

\begin{document}

\begin{titlepage}

\topmargin -15mm

\vskip 10mm

\centerline{ \LARGE\bf Effects of Disorder in Two-Dimensional }
\vskip 2mm
\centerline{ \LARGE\bf Quantum Antiferromagnets}

    \vskip 2.0cm

    \centerline{\sc C.M.S.Concei\c c\~ao and E.C.Marino }

     \vskip 0.6cm
     
    \centerline{\it Instituto de F\'\i sica }
    \centerline{\it Universidade Federal do Rio de Janeiro }
    \centerline{\it Cx.P. 68528}
    \centerline{\it Rio de Janeiro RJ 21941-972}
    \centerline{\it Brazil}

\vskip 1.0cm

\begin{abstract} 
 
We study the effects of disorder in two-dimensional quantum antiferromagnets on a square
lattice, within the nonlinear sigma model approach, by using of a random
distribution of spin stiffnesses or zero-temperature-spin-gaps, respectively, in the
renormalized classical and quantum disordered phases. The quenched staggered 
magnetic susceptibility at low temperatures is evaluated in each case. The 
large distance behavior of the quenched spin correlation function is also obtained in
the quantum disordered phase. Disorder is shown to introduce a change from exponential
to power-law decay in these functions, 
indicating that the spin excitations become gapless, in spite of
the fact that the system is in a disorder state. A comparison is made with the dual
behavior of skyrmion topological excitations in the renormalized classical phase.
			      
\end{abstract}

\vskip 0.5cm


\vskip 0.5cm


\end{titlepage}

\hoffset= -10mm

\leftmargin 23mm

\topmargin -8mm
\hsize 153mm
 
\baselineskip 7mm
\setcounter{page}{2}

\section{Introduction}

The physics of two-dimensional quantum antiferromagnetic systems has been attracting a lot 
of interest, to a great extent due to 
the important role played by them in the high-Tc
superconducting cuprates. The continuum description of these systems in terms of a Nonlinear
Sigma Model (NLSM), in particular, has proved to be very convenient in order to extract 
relevant physical information such as the phase structure, order parameters, magnetic
susceptibilities, correlation functions and critical exponents, among others 
\cite{ha,chn,qaf,cs,st}.

In a recent paper \cite{em1}, the effect of disorder has been investigated in the 
renormalized classical, N\'eel ordered, phase of the NLSM. This was done through the 
introduction of a random continuum distribution for the spin stiffness, which is the 
relevant control parameter in this phase. Quenched averages were then used to evaluate
the effects of disorder on the correlation functions of quantum topological excitations 
(skyrmions). An interesting consequence was found concerning the energy gap of these
excitations. In the N\'eel phase, without disorder, 
skyrmions have a finite energy gap, proportional to the spin stiffness ($\rho_s$), 
which is the square of the staggered magnetization \cite{em2}. 
This fact usually enables one to use
the skyrmion energy gap as a reliable order parameter for the N\'eel phase.
When disorder is introduced, in such a way that magnetic dilution is not exponentially 
supressed, however, one finds that the 
skyrmion energy gap vanishes, in spite of the fact that the quenched staggered magnetization
is different from zero \cite{em1}. This means that in certain situations one cannot use
the skyrmion gap as a suitable order parameter for the AF order.

In the present work, we analyze the effects of disorder on the spin correlation function
in the quantum paramagnetic (quantum disordered) phase of the NLSM and also on the 
staggered magnetic susceptibiliy, both in the N\'eel and quantum paramagnetic phases.
For this purpose, we exploit the order-disorder duality relation existing between the spin
excitations and the quantum skyrmion topological excitations \cite{em2}. Indeed, 
in the ordered N\'eel phase, skyrmions possess a nonzero energy gap while spin-waves 
(magnons) are gapless whereas in the quantum paramagnetic phase the former are gapless 
and the latter acquire a nonzero gap. The relevant control parameter of the system 
in the quantum paramagnetic phase is the spin gap at zero temperature, $\Delta_s$. 
Consequently, in order to
describe the presence of disorder in the system in the quantum paramagnetic phase, 
we shall introduce a random distribution for the 
spin gap and evaluate quenched averages of relevant
quantities. A further
justification for this way of introducing disorder in the quantum paramagnetic phase
comes from the way $\rho_s$ and $\Delta_s$ are expressed in terms of the coupling constant
of the NLSM ($g_0$). Indeed, we have  
\be
\rho_s = \fr{1}{g_0} - \fr{1}{g_c} \ \ ; \ \ \Delta_s = 8\pi \lef ( \fr{1}{g_c} 
- \fr{1}{g_0} \ri ) ,
\label{1}
\ee
where $g_c$ is the critical coupling for the quantum phase transition 
separating the N\'eel and the quantum paramagnetic phases \cite{chn}. 
$g_0$ is directly related to the exchange coupling of 
the original spin system and therefore it is natural to expect that the effects of 
disorder existing in the latter should manifest through $g_0$ in the continuum version.
Consequently, in view of (\ref{1}), corresponding to a random distribution of $\rho_s$
in the N\'eel phase, we must have a similar distribution of $\Delta_s$ in the quantum 
paramagnetic phase. We shall use, both for the spin stiffness and spin gap, 
respectively in each phase, the distribution \cite{em1}
\be
P[\alpha] =
 \lef \{  \begin{array}{c}
\fr{1}{N}  \alpha^{\nu -1}
 e^{- \fr{(\alpha - \alpha_s)^2}{2 \sigma^2}}\ \
 \alpha \geq 0  \\    \\
 0 \ \ \ \ \ \ \ \ \ \ \ \ \ \ \ \ \ \ \ \ \ \ \  \alpha < 0
          \end{array} \ri .
\label{2}
\ee
where $\nu > 0$ and 
\be
N = \sigma^\nu \Gamma(\nu) D_{-\nu}\lef (-\fr{\alpha_s}{\sigma}\ri )
e^{-\fr{\alpha_s^2}{4\sigma^2}}
\label{3}
\ee
where $D_{-\nu}(x)$ is a parabolic cylinder function. In (\ref{2}), $\alpha$ stands either
for the spin stiffness $\rho$, in the ordered N\'eel phase ($\alpha_s \equiv \rho_s$), 
or for the spin gap $\Delta$, in the quantum paramagnetic phase ($\alpha_s \equiv \Delta_s$). 
Notice that the distribution vanishes for negative values of the argument. This is a natural 
choice since the control parameters $\rho$ and $\Delta$ are not defined in this case. 
Ferromagnetic couplings, in particular, are not described by the distribution above.

In the case of the high-Tc cuprates, which are prototypes of 2D quantum antiferromagnets,
typical values for the parameters are \cite{ak,mm}:
$\rho_s \simeq 10^{-1} eV$; $\Delta_s \simeq 10^{-3} eV$. 
We assume $\sigma << \rho_s$, and $\sigma << \Delta_s$, respectively, in each phase,
in order to ensure that the $\rho$ and $\Delta$ configurations are slowly varying and
the continuum limit can be safely taken.
Hence, typical values for the variance would be $\sigma \simeq 10^{-3}eV$ and 
$\sigma \simeq 10^{-5}eV$, respectively,
in the ordered N\'eel and quantum paramagnetic phases. Also, for a sample of
dimension $L \simeq 1mm$ we have, in both phases, the relation (to be used later on)
\be
 \lef ( \fr{L}{\hbar c}\ri ) \sigma >> \lef (\fr{\alpha_s}{\sigma} \ri ) >> 1 , 
\label{4}
\ee
where $c$ is the spin-wave velocity, the characteristic velocity of the system. The typical
value for the ordered phase is $\hbar c \simeq 1 eV \AA $ \cite{ak} and we assume 
$\hbar c \simeq 10^{-2} eV \AA $ in the quantum paramagnetic phase.

\section{Renormalized Classical Region}

Let us investigate here the effects of disorder on the homogeneous, static 
staggered susceptibility for $g_0 < g_c$, or $\rho_s > 0$. Within the CP$^1$
formulation of the NLSM, this is given by \cite{st}
\be
\chi(T) = \lim_{|\vec k|, \omega \rightarrow 0} \chi(\omega, \vec k, T) =
\frac{T}{4 m^2(T)}   \ ,
\label{5}
\ee
where $\chi(\omega, \vec k, T)$ is the vacuum polarization scalar for the CP$^1$
constraint field and $m(T)$ is the spin gap. For $T << \rho_s $, we have
$m(T) = T e^{-\frac{2\pi \rho_s}{T}}$ and 
\be
\chi(T) = \frac{e^\frac{4 \pi \rho_s}{T}}{4T} 
\stackrel{T \rightarrow 0} {\longrightarrow} \infty
\label{6}
\ee
The spin gap vanishes at $T=0$ and the susceptibility diverges, implying the occurence 
of an ordered N\'eel state at zero temperature.

In the presence of disorder, the quenched susceptibility will be given by
\be
\bar\chi(T) = \int_0^\infty d\rho P[\rho] \chi(\rho,T) ,
\label{7}
\ee
where $P[\rho]$ is given by (\ref{2}). The condition $\sigma << \rho_s$ enables us to use
the expression (\ref{6}) in (\ref{7}), obtaining, for $T << \sigma << \rho_s$
\be
\bar\chi(T) =  \frac{1}{4 T^\nu}  
\lef [ \frac{4 \pi \sigma^2}{\rho_s}  +  T   \ri ]^{\nu -1}
\exp \lef [ \frac{8 \pi^2 \sigma^2}{T^2} + \frac{4 \pi \rho_s}{T} \ri ]
 ,
\label{8}
\ee
We see that we have 
$\bar\chi(T) \stackrel{T \rightarrow 0} {\longrightarrow} \infty$. This means that
even after the inclusion of disorder the ordered N\'eel state persists for 
$\rho_s > 0$, a result that confirms, from the point of view of the
susceptibility, the observation made in \cite{em1} by looking
directly at the quenched magnetization. It has been shown, nevertheless, that
the quantum skyrmion quenched correlation function changes its behavior from exponential to 
power-law decay at large distances as a consequence of the presence of disorder
in this phase \cite{em1}.

\section{Quantum Paramagnetic Phase}

We now consider the effects of disorder on the susceptibility in the
quantum paramagnetic phase, characterized by $g_0 > g_c$, 
or $\Delta_s > 0$. $\chi(T)$ is still given by (\ref{5}) but the spin gap
is now given, for $T << \Delta_s$, by \cite{cs} 
\be
m(T) = \Delta_s + 2T e^{-\frac{\Delta_s}{T}}
\label{9}
\ee 
and no longer vanishes at zero temperature. The susceptibility is now given by
\be
\chi(T) = \frac{T}{4 \Delta_s^2(T)}\lef (1 - 4 \frac{T}{\Delta_s} e^{-\frac{\Delta_s}{T}}
 \ri )
\label{10}
\ee
and we see that $\chi(T) \stackrel{T \rightarrow 0} {\longrightarrow} 0$, implying 
the absence of an ordered state at $T=0$ in this phase.

We now introduce disorder in the system and consider the quenched susceptibility
\be
\bar\chi(T) = \int_0^\infty d\Delta P[\Delta] \chi(\Delta,T) ,
\label{11}
\ee
where $P[\Delta]$, the random distribution of $\Delta$'s, is given by (\ref{2}) and
$\chi(\Delta,T)$, by (\ref{10}). Observe that the condition  $\sigma << \Delta_s$
allows us to introduce the $T/\Delta_s$-expanded expression (\ref{10}) inside the 
integral in (\ref{11}). After performing the $\Delta$ integration we get, for 
$T << \Delta_s$
\be
\bar\chi(T) = \frac{T}{4 \Delta_s^2} - \frac{T^2}{\Delta_s^3} 
\exp \lef[- \frac{\Delta_s}{T}\lef ( 1 - \frac{\sigma^2}{2 T \Delta_s} \ri ) \ri ]
\lef [ 1 - (\nu -4) \frac{\sigma^2}{T \Delta_s}  \ri ] .
\label{12}
\ee
We see that $\bar\chi(T) \stackrel{T \rightarrow 0} {\longrightarrow} 0$, implying that
the inclusion of disorder, described by (\ref{2}) does not change the fact that 
the system is in a disordered ground state, at $T = 0$, for 
$\Delta_s > 0$ as one should expect. Notice,
however, that the subleading behavior of the susceptibility is modified by 
the presence of disorder.
Observe, on the other hand, that disorder does not change the critical exponent for 
the quantum phase transition occurring for $g \rightarrow g_c$, at $T \rightarrow 0$.
In any case, we have the susceptibility behaving as $(g - g_c)^{-2}$ at the critical 
point.

Let us investigate now the large distance behavior of the spin correlation function
$<S_z (\vec x, \tau)S_z (\vec 0, \tau)>$. This is proportional to \cite{st}
\be
\chi(\vec x) = \int \frac{d^2k}{(2\pi)^2} 
\chi(\omega = 0, \vec k, T) e^{i \vec k \cdot \vec x}
\stackrel{ |\vec x| \rightarrow \infty} {\longrightarrow} e^{-2 m(T)|\vec x|} ,
\label{13}
\ee
where $m(T)$ is given by (\ref{9}) in the quantum paramagnetic phase. 

We are interested 
in the zero temperature, large distance behavior of the quenched spin correlation function, 
hence we consider
\be
\bar\chi(\vec x) = \int_0^\infty d\Delta P[\Delta] \chi(\vec x, \Delta, T=0)
\stackrel{ |\vec x| \rightarrow \infty} {\longrightarrow} 
\int_0^\infty d\Delta P[\Delta]  e^{-2 \Delta |\vec x|} . 
\label{14}
\ee
Performing the $\Delta$ integration, we get, for $\sigma << \Delta_s$,
\be
\bar\chi(\vec x) =  \frac{\Gamma (\nu) }{\sqrt{2\pi}} 
\lef ( \frac{\sigma}{\Delta_s} \ri )^{\nu -1} e^{-\Delta_s^2/2\sigma^2}
\exp \lef [ \lef (\sigma |\vec x| -\frac{\Delta_s}{2 \sigma}  \ri )^2  \ri ]
D_{-\nu}\lef (2 \sigma |\vec x| -\frac{\Delta_s}{ \sigma}  \ri ) .
\label{15}
\ee
In order to obtain the large distance behavior of (\ref{15}), we use condition
(\ref{4}) and the asymptotic behavior of the parabolic cylinder functions \cite{gr}.
The result is
\be
\bar\chi(\vec x) =  \stackrel{ |\vec x| \rightarrow \infty} {\longrightarrow} 
\frac{\Gamma (\nu) }{2^\nu \sqrt{2\pi}} 
\lef ( \frac{\Delta_s^{1 - \nu }}{\sigma} \ri )  e^{-\Delta_s^2/2\sigma^2}
\frac{1}{|\vec x|^\nu} 
\lef [1 + \nu \frac{\Delta_s}{2 \sigma^2 |\vec x|}    \ri ]
\label{16}
\ee
The power-law large distance behavior of the quenched spin correlation function indicates 
that, in the presence of disorder, associated to the distribution of $\Delta$'s
given by (\ref{2}), the spin excitations become gapless for
$\Delta_s > 0$ and $T=0$. This happens, in spite of the fact that the system is in a
disordered state, as we have seen before. In such sates,
the spin excitations usually have a gap and therefore, the introduction of disorder
completely changes the scenario in this case. As we have mentioned something similar
happened with the skyrmion topological excitations in the renormalized classical
region thereby clearly exposing the duality existing between skyrmions and
spin excitations.

\section{Concluding Remarks}

We have considered the inclusion of disorder in a two-dimensional quantum antiferromagnet, 
both in the renormalized classical and quantum paramagnetic phases, at low temperatures. 
Staggered magnetic susceptibilities and spin correlation functions have been studied in the 
quenched regime. The behavior of these indicate
that the nature of the ground state is not modified but, nevertheless, the properties 
of the basic excitations are deeply changed. Spin excitations, become gapless in the 
quantum paramagnetic phase, thereby exhibiting a dual character with respect 
to skyrmion quantum excitations, which become gapless, in the presence of disorder,
within the renormalized classical region \cite{em1}.

In this work, we described the presence of disorder by means of a continuuum 
random distribution of the relevant control parameter in each phase, 
either the spin stiffness or the zero-temperature-spin-gap. It would be interesting to
investigate the effect of a discrete distribution thereof, which could be
introduced by delta functions. Also, the types of disorder in which ferromagnetic
couplings would be allowed would be worth studying and could have interesting 
consequences for models of spin-glasses. We are presently considering these 
possibilties.

\leftline{\bf Acknowledgements}

This work has been supported in part by CNPq, FAPERJ
and PRONEX-66.2002/1998-9. CMSC has been supported by CAPES and FAPERJ and
ECM has been partially supported by CNPq.


\begin{thebibliography}{99}


\bibitem{ha} F.D.M.Haldane, {\it Phys. Rev. Lett.} {\bf 50}, 1153 (1983);
F.D.M.Haldane, {\it Phys. Lett.}, {\bf A93} 464 (1983)

\bibitem{chn} S.Chakravarty, B.Halperin and D.Nelson,
{\it Phys. Rev.} {\bf B39}, 2344 (1989)

\bibitem{qaf} N.Read and S.Sachdev,
{\it Phys. Rev.} {\bf B42}, 4568 (1990);
A.V.Chubukov, S.Sachdev and J.Ye, {\it Phys. Rev.} {\bf B49}, 11919 (1994);
V.Y.Irkhin and A.A.Katanin, {\it Phys. Rev.} {\bf B55}, 12318 (1997)

\bibitem{cs} A.V.Chubukov and O.Starykh, {\it Phys. Rev.} {\bf B52}, 440 (1995)

\bibitem{st} O.Starykh, {\it Phys. Rev.} {\bf B50}, 16428 (1994)

\bibitem{em1} E.C.Marino, {\it Phys. Rev.} {\bf B65}, 054418 (2002) 

\bibitem{em2} E.C.Marino, {\it Phys. Rev.} {\bf B61}, 1588 (2000) 

\bibitem{ak} A.P.Kampf, {\it Phys. Rep.} {\bf 249}, 219 (1994)

\bibitem{mm} E.C.Marino and M.B.Silva Neto, {\it Phys. Rev.} {\bf B66}, 224512 (2002) 

\bibitem{gr} I.S.Gradshteyn and I.M.Ryzhik, {``Table of Integrals, Series
and Products''}, Academic Press, New York, 1980








 
\end{thebibliography}
\end{document}